# Impact of refractive index heterogeneity on stimulated Brillouin scattering microscopy: a quantitative analysis

Meng Xu,[1,2,†] Zixuan Du,[1,2,†] Yun Qi,[1,2,*] Jinrui Zhang,[1] Shuai Yao,[1,2] Robert Prevedel,[3] and Fan Yang,[1,2,*]

[1] Shanghai Institute of Optics and Fine Mechanics, Chinese Academy of Sciences, Shanghai, China
[2] University of Chinese Academy of Sciences, Beijing, China
[3] European Molecular Biology Laboratory, Heidelberg, Germany

[†]These authors contributed equally to this work.
*qiyun@siom.ac.cn;

*yang@siom.ac.cn

**Stimulated Brillouin scattering (SBS) microscopy enables label-free biomechanical imaging, with Brillouin gain serving as a critical contrast parameter for quantitative analysis. However, the influence of sample-induced refractive index (RI) heterogeneity on gain measurements remains poorly understood. Here, we quantitatively investigate, how RI mismatch affects SBS microscopy using finite element simulations and experiments on a phantom sample comprising polydimethylsiloxane beads embedded in agarose gel. We demonstrate that RI heterogeneity induces focal field distortion that reduce pump-probe beam overlap, resulting in attenuated Brillouin gain and degraded shift precision at material interfaces. Crucially, we establish that fiber-coupling efficiency—commonly used for system alignment—cannot serve as a linear proxy for Brillouin gain due to its heightened sensitivity to focal field distortion.**

Brillouin microscopy enables biomechanical imaging in an all-optical, non-contact and three-dimensional manner with high spatial resolution [1-3]. Although stimulated Brillouin scattering (SBS) microscopy requires two-end optical access, it offers higher spectral resolution than spontaneous Brillouin microscopy [4]. The use of pulsed pump and probe beams in SBS microscopy can significantly enhance performance. Specifically, pulse-chopping approaches have reduced optical power by a factor of ten, enabling low-phototoxicity biomechanical imaging for fragile samples [5,6]. More recently, a pulsed-laser scheme employing high peak power and low duty cycle has enhanced pixel acquisition speed by two orders of magnitude, achieving a 200 μs pixel dwell time with full spectral information [7]. An additional advantage of SBS microscopy is that it potentially provides three contrast parameters, including Brillouin shift, linewidth and gain, compared to the two typically available in spontaneous methods [4]. Unlike spontaneous Brillouin scattering, where changes in signal intensity cannot be easily separated from pump attenuation effects in absorbing or heterogeneous media, SBS enables built-in normalization by simultaneously monitoring the DC baseline of the transmitted probe and the AC component of the SBS signal, thereby allowing more reliable extraction of the intrinsic Brillouin gain. By combining these three parameters with the sample's refractive index, one can, in principle, derive the dry mass density and longitudinal modulus without relying on assumptions regarding the ratio of mass density to refractive index [4,8]. Therefore, the Brillouin gain of the sample is critical. However, while gain is known to be affected by the overlap between focused counter-propagating pump and probe beams, a quantitative analysis of this effect is currently lacking.

In this work, we quantitatively investigate the impact of heterogeneous refractive index (RI) distributions, which induce focal field distortion, on the Brillouin gain. We focus on the critical axial dimension, where the impact of focal field distortions on beam overlap is most pronounced. To this end, we simulate a model system comprising a polydimethylsiloxane (PDMS) bead embedded in 1% (w/v) agarose gel to examine how the heterogeneous RI influences the focal fields distributions of pump and probe beams, thereby degrading the pump-probe overlap and consequently altering the Brillouin gain and other spectral parameters. To verify our theoretical simulations, we conducted axial SBS imaging experiments on microbeads under analogous conditions. The experimental findings show strong agreement with the simulation data, advancing our understanding of the mechanisms behind SBS signal generation and providing a quantitative basis for evaluating the accuracy of Brillouin gain measurements in heterogeneous environments. Furthermore, we conclude that the coupling efficiency $\eta_c$ —although used during optical alignment—is more sensitive than Brillouin gain and therefore cannot serve as a linear predictor of the Brillouin gain.

**Theoretical Model.** To elucidate the mechanisms underlying the experimentally observed attenuation of Brillouin gain, it is necessary to revisit the fundamental of SBS. SBS is a third-order nonlinear optical effect that originates from the interaction between light waves and acoustic phonons within a medium. The typical configuration in our point-scanning SBS microscopy employs two collinear, counter-propagating beams: a pump beam at frequency $\omega_P$ and a probe (Stokes side) beam at frequency $\omega_S$. The experimental setup is based on a point-scanning pulsed SBS microscopy system, the configuration of which is similar to that reported in Ref. [5] (Supplementary Note 1). According to SBS theory, the energy transfer between the counter-propagating beams is modeled by the coupled Equations (1) and (2), which describe the intensity changes of the pump ($I_P$) and probe ($I_S$) beams along the z axis [9]:

$$dI_S(x,y,z) = g(\Omega)I_P(x,y,z)I_S(x,y,z)dz \quad (1)$$

$$dI_P(x,y,z) = -g(\Omega)I_P(x,y,z)I_S(x,y,z)dz \quad (2)$$

where $g(\Omega)$ is the Brillouin gain factor, exhibiting a Lorentzian distribution in the frequency domain as shown in Equation (3).

$$g(\Omega) = g_0 \frac{(\Gamma_B/2)^2}{(\Omega - \Omega_B)^2 + (\Gamma_B/2)^2} \quad (3)$$

In this expression, $g_0$ is the Brillouin line-center gain factor. $\Omega = \omega_P - \omega_S$ is the angular frequency difference between the pump and probe beam. $\Gamma_B$ is the linewidth of the Brillouin gain spectrum, typically defined as the full width at half maximum (FWHM) of the Brillouin gain factor. $\Omega_B$ is the Brillouin frequency shift, defined as the frequency difference at which the Brillouin gain factor reaches its maximum value.

Experimentally, our setup detects the integrated power of the transmitted probe beam by a photodetector. Therefore, to theoretically model the probe power evolution ($P_S(\Omega, z)$), the local gain intensity must be integrated over the entire interaction volume:

$$\Delta P_S(\Omega) = \iiint_V g(\Omega)I_P(x,y,z)I_S(x,y,z)dV \quad (4)$$

The small gain approximation (~$10^{-6}$) has been adopted and the beam intensities are treated as constant after interaction. Equation (4) shows that the contribution of each spatial location to $\Delta P_S$ is weighted by the local product of the pump and probe intensities, and the volumetric integral determines the global SBS signal strength. In our experiment, the detected SBS signal spectrum is then fitted with a Lorentzian function, and the peak amplitude is extracted as the SBS signal amplitude ($A_{SBS}$) which is linearly proportional to the Brillouin gain and $g_0$ [4-7].

**Modeling the Spatial Overlap of Pump and Probe Beams in SBS Microscopy.** To further understand how RI heterogeneity affects the overlap between pump and probe beams, we modeled a system consisting of a ~8-μm-diameter PDMS bead (n ≈ 1.41) embedded in 1% (w/v) agarose gel (n ≈ 1.33). The respective field distributions for a forward-propagating pump beam and a counter-propagating probe beam which focused at the same spatial location are both calculated using the finite element method in COMSOL Multiphysics (Supplementary Note 2) at a wavelength of 780 nm with linear polarization (see Supplementary Note 3 for justification of the polarization choice). Figure. 1(a) schematically illustrates the spatial configuration of the pump focus, probe focus, and bead. The intensity distributions of the pump and probe beams within the region highlighted by the solid red square in Fig. 1(a) are shown in Fig. 1(b) and 1(c), respectively. High-quality focus fields are maintained despite the bead introducing a secondary focusing effect. Thus, as shown in Fig. 1(d), a strong spatial overlap is obtained, indicating a high Brillouin gain can be retrieved. Figure. 1(e) shows another representative configuration, in which the nominal focal point is located just outside the upper-left boundary of the bead. In this case, a substantial portion of the focused pump beam is distorted by the bead and deflected from its original path (Fig. 1(f)), whereas the counter-propagating probe beam is unaffected and remains well focused at the intended location (Fig. 1(g)). As a result, the spatial overlap between the pump and probe beams is markedly reduced, leading to a weakened SBS signal, as illustrated in Fig. 1(h). Therefore, RI heterogeneity must be carefully considered when an accurate determination of the absolute Brillouin gain is required.

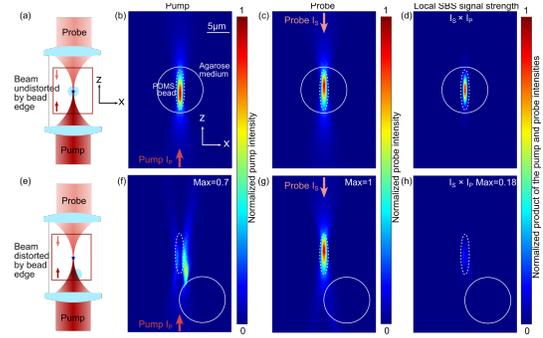

**Fig. 1.** The effect of RI distribution (PDMS bead) on actual focal position and SBS signal strength. (a, e) Schematics of the collinear pump and probe beams with their nominal focal point (blue star) positioned at (a) the bead center and (e) just outside the upper-left boundary of the bead, respectively. (b–d) Simulated intensity distributions within the red-boxed region shown in (a): (b) pump beam, (c) probe beam, and (d) the corresponding local SBS signal strength. (f–h) Simulated intensity distributions within the red-boxed region shown in (e): (f) pump beam, (g) probe beam, and (h) the corresponding local SBS signal strength. The white solid circle denotes the physical boundary of the PDMS bead. The white dashed elliptical contour represents the nominal focal region of the beams in the absence of the bead. All the data shown in Fig. 1(b–d) were normalized to the maximum intensity of each map and Figs. 1(f–h) for which the maximum values were marked at the top-right were respectively normalized relative to Figs. 1(b–d).

**Simulated Brillouin gain map and experimental verification.** To more comprehensively evaluate the influence of the phantom on the Brillouin gain, we simulated the entire axial $A_{SBS}$ image by mimicking a cross-sectional raster scan used in conventional imaging. To make the computational burden manageable, reducing the simulation time from months to under two weeks, we employed a manually defined variable step size scheme (1 μm coarse and 0.25 μm fine steps based on region) and subsequently resampled the entire dataset to a uniform 0.25 μm resolution using thin-plate spline interpolation [10]. For computational efficiency, a reduced sampling rate was applied in regions where the Brillouin gain varies slowly (e.g., in agarose gel regions far from the bead). Specifically, the $A_{SBS}$ of agarose gel and PDMS bead are calculated by integrating the product $\kappa I_P(x,y,z)I_S(x,y,z)$ over their respective spatially occupied regions, where material-specific factors $\kappa = max(A_{SBS})/max(\iiint I_P I_S dV)$ accounts for both the material gain $g(\Omega_B)$, beam powers, and system responses. Figure 2 shows the simulated spatial distributions of $A_{SBS}$ with the pump-probe frequency shift tuned to the Brillouin shift of the PDMS bead (Fig. 2(a)) and agarose (Fig. 2(c)), both of which agree well with the experimentally obtained ASBS distributions (Fig. 2(b) and 2(d)), confirming the accuracy of the simulation. In addition, a clear $A_{SBS}$ reduction is observed in the agarose gel regions adjacent to the bead's four quadrants (marked by black boxes), consistent with the decreased spatial overlap of the pump and probe beams mentioned previously. This phenomenon is generally widespread and has been observed in the imaging of notochord of zebrafish larvae [7]. It should be emphasized that previous works only observed the axial gain distribution, this paper has quantitatively analyzed the decrease of the gain in the SBS image for the first time to our knowledge. Notably, the stronger water signal appearing within the bead region in the simulated data, relative to the experimental measurements, can be attributed to the slightly reduced axial resolution implemented in the simulation. This interpretation is further corroborated by the more evident axial elongation artifacts in the simulated bead image (Fig. 2(a)), as compared with the experimental image (Fig. 2(b)).

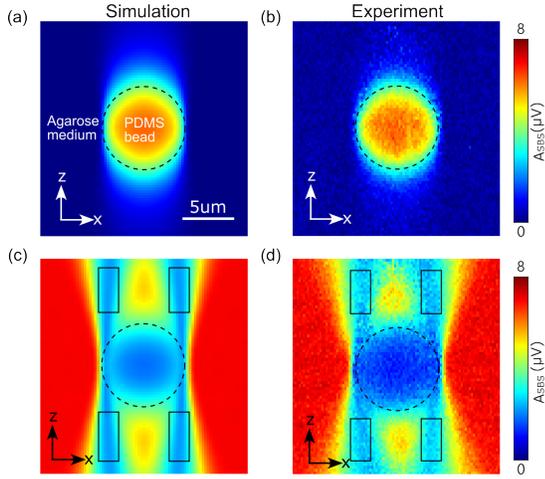

**Fig. 2.** (a) and (b) simulated and experimental results of $A_{SBS}$ distribution of PDMS bead along the XZ plane, respectively. (c) and (d) simulated and experimental results of $A_{SBS}$ distribution of 1% agarose gel along the XZ plane, respectively. In the experiment, double peak fitting is applied to retrieve the $A_{SBS}$ for each substance. Black dashed circles: approximate bead position. Solid black boxes: regions of $A_{SBS}$ attenuation in agarose gel.

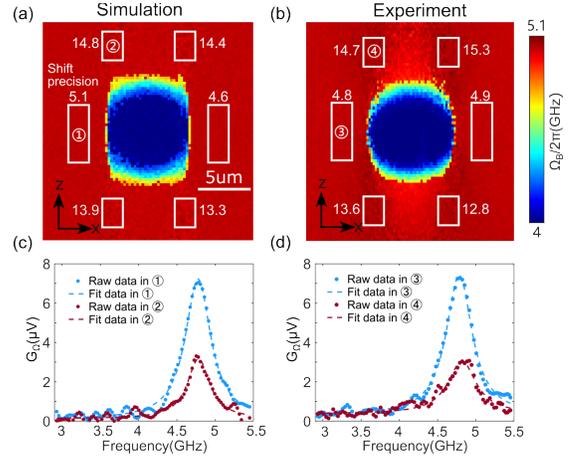

**Fig. 3.** (a) and (b) simulated and experimental results using the single peak fitting of the Brillouin shift distribution along the XZ plane, respectively. Values beside the white solid boxes indicate the standard deviation of Brillouin shifts within that region, quantifying fitting precision. (c) and (d) simulated and experimental SBS spectra in region 1/2 and 3/4, respectively.

**Effect of pump–probe overlap on single-peak-fitting Brillouin shift.** As the measured spectrum represents a superposition of the spectra from all constituents within the confocal volume, synthetic SBS spectra were generated by quantitatively summing the spectra of PDMS (Brillouin shift/linewidth: 3.98 GHz/880 MHz) and 1% agarose gel (5.03 GHz/410 MHz), weighted by their respective $A_{SBS}$ maps as shown in Figs. 2(a) and 2(c). In the SBS experiments, the noise is dominated by electronic noise from the detection system. Accordingly, Gaussian noise with a fixed standard deviation, filtered by an 800 Hz low-pass filter, was added to all synthetic spectra to emulate the experimental conditions. The resulting simulated spectra were then fitted using a single-Lorentzian model to extract the Brillouin spectral parameters, consistent with the analysis employed in [4–7], yielding an apparent Brillouin shift transition between agarose gel and the bead at the boundary. Excellent agreement between the simulated and experimental Brillouin shift distributions is observed in Figs. 3(a) and 3(b).

In addition, a clear degradation in shift precision—defined as the standard deviation of the fitted Brillouin shift—is observed in the four diagonal regions surrounding the bead interface, as shown in Figs. 3(a) and 3(b). This degradation arises from the reduced Brillouin gain caused by poor pump–probe overlap, while the noise level remains constant (Figs. 3(c) and 3(d)). These results indicate that although the absolute Brillouin shift remains largely unaffected in regions with diminished overlap, the overall imaging quality is substantially compromised. Similar effects have been experimentally measured previously in longitudinal Brillouin sectioning of cells [5].

It is worth noting that the rectangular appearance of the bead in the simulated image arises from insufficient sampling at the boundary, where the optical field varies rapidly. In addition, the regions above and below the bead in Fig. 3(b) exhibit slightly reduced Brillouin frequency shifts. This effect can be attributed to additional focusing of the pump or probe beam by the bead, which introduces higher spatial-frequency components and leads to angular broadening of the SBS spectrum, thereby reducing the fitted Brillouin shift [11]. In the simulations, however, the spectra are generated by direct spectral synthesis rather than being retrieved from a full physical interaction that includes this angular-broadening effect. Thus, this reduction in Brillouin shift is not observed in the simulated results and is beyond the scope of the present study. Overall, these results demonstrate that reduced pump–probe overlap weakens the SBS signal and directly accounts for the degraded signal-to-noise ratio and fitting precision commonly observed at axial biological boundaries in previous Brillouin imaging studies [5,7].

**Insufficiency of using norm. $\eta_c$ as a proxy for Brillouin gain.** In previous studies, pump–probe overlap was typically optimized by maximizing the $\eta_c$ of one beam into the opposing fiber port, which effectively aligns the two foci within the sample [5,7]. This practice implicitly assumes that $\eta_c$ serves as a reliable proxy for the Brillouin gain. If such a linear relationship existed, $\eta_c$ could potentially be used as a convenient calibration metric to correct gain estimation errors arising from RI mismatch. Motivated by this idea, we investigated the relationship between $\eta_c$ and $A_{SBS}$ through experiments and numerical simulations (see Supplementary Note 4 for details).

Because full 3D modeling over the large interaction volume is computationally prohibitive, we use a simplified 2D model to capture the essential physics for our analysis. We analyzed the optical fields at two key locations: the focal region, which determines the local $A_{SBS}$, and the receiving fiber entrance, which dictates the $\eta_c$. Figure 4(a) shows the reference case without bead interaction, where a collimated pump beam is focused into the homogeneous medium and coupled back into the probe fiber along a symmetric path, yielding $\eta_c$ ~97% in the absence of RI-mismatch perturbation (Fig. 4(b)). Figure 4(c) shows the beam focused near the bead edge; as seen in Fig. 4(d), the induced beam distortion is amplified during propagation, severely degrading mode matching at the fiber entrance and reducing the $\eta_c$ to ~5%, while the medium's $A_{SBS}$ drops only to ~25% of its peak value.

Figures. 4(e) and 4(f) show the experimental and simulated norm. $\eta_c$ distributions along the XZ plane centered on bead respectively. Both exhibit the norm. $\eta_c$ attenuation in the marked regions, occurring at analogous spatial positions as those observed in Figs. 2(c) and 2(d). However, the decrease of norm. $\eta_c$ is notably sharper than that of norm. $A_{SBS}$ as Figs. 4(b) and 4(d) shown. This divergence between norm. $A_{SBS}$ and norm. $\eta_c$ within the region indicated by the white mask is quantified as a relative difference in Supplementary Fig. S5. This contrast, spanning both bead-affected and unaffected regions, effectively demonstrates the fundamental insufficiency of using norm. $\eta_c$ as a proxy for Brillouin gain (see Supplementary Note 5 for quantitative analysis).

Notably, the measured distribution shows a slight asymmetry between the bead's upper and lower regions, likely caused by minor optical-axis misalignments during assembly that lead to small differences in beam convergence or divergence after transmission and accumulate over long propagation distances. Here, the peak values in Figs. 4(a) and 4(b) are self-normalized to unity to facilitate intuitive comparison between two cases, since under simulation conditions with negligible system aberrations the theoretical maximum $\eta_c$ for the pump beam focused on the homogeneous medium can approach 97%, whereas in practical systems optical aberrations and assembly tolerances typically limit this peak efficiency to ~50% in real-world systems.

Collectively, although maximizing $\eta_c$ is the protocol for system alignment [5,7], one might intuitively propose extending its use as a linear calibration factor to compensate for signal loss from reduced pump-probe overlap. However, our findings establish the insufficiency of this approach. Since beam distortion amplifies during propagation, $\eta_c$ is far more sensitive to focal field distortion than Brillouin gain, and thus cannot serve as a reliable proxy for the latter.

In summary, we have presented a comprehensive investigation into how sample-induced RI mismatch quantitatively affects SBS microscopy. Our theoretical and experimental results show excellent agreement, demonstrating that RI heterogeneity distorts the focal field and reduces beam overlap, leading to attenuated Brillouin gain and decreased Brillouin shift-fitting precision. Crucially, we demonstrate that using norm. $\eta_c$ as a proxy for Brillouin gain is insufficient. These findings may provide essential insight for developing accurate calibration strategies in quantitative SBS imaging. Employing a confocal pinhole before photodetector to normalize the true spatial intensity overlap could be a potential solution to calibrate the gain. Numerically, extending beyond our proof-of-concept model, efficient algorithms could calculate location-dependent correction factors for complex scenarios (e.g., multiple scatterers) by taking the reconstructed RI (via optical diffraction tomography) map and the aberration-inclusive incident field as inputs. Alternatively, adaptive optics could physically correct these distortions to restore focal overlap, ensuring high-fidelity imaging.

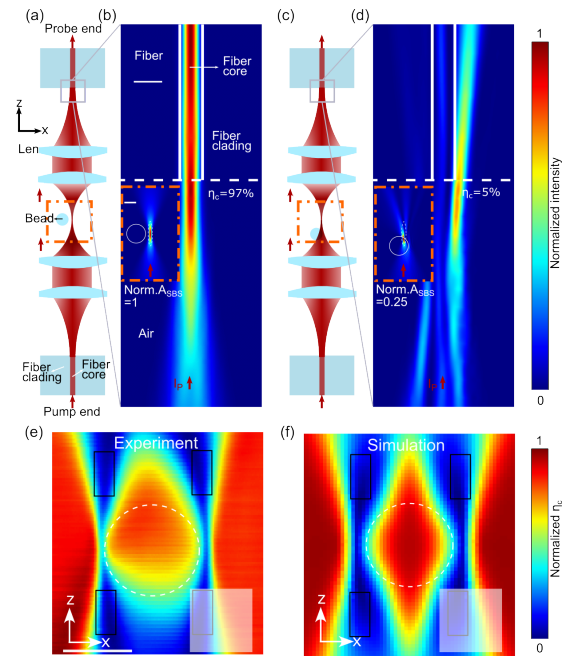

**Fig. 4.** (a), (c) Schematics of the beam configuration (a) laterally displaced from the bead (unperturbed) and (c) focused at the bead edge. (b), (d) Corresponding intensity distributions at the fiber coupling end; insets show the intensity pump focal spots. Intensities in (b) are self-normalized, while (d) is normalized to the peaks in (d). White dashed circles: approximate bead position. Solid black boxes: regions of $\eta_c$ attenuation. (e), (f) Measured and simulated XZ norm. $\eta_c$ maps. Scale bars: 5 μm. White mask: region of interest spanning both bead-affected and unaffected regions for the error analysis presented in Fig. S5.

**Funding.** Ministry of Science and Technology of the People's Republic of China (2024YFA1612102); Chinese Academy of Sciences (XDB0650000); Shanghai Institute of Optics and Fine Mechanics; European Molecular Biology Laboratory; European Research Council (CoG, no. 864027).

**Disclosures.** The authors declare no conflicts of interest.

**Data availability.** Data underlying the results presented in this paper are not publicly available at this time but may be obtained from the authors upon reasonable request.

**Supplemental document.** See Supplement 1 for supporting content.

# SUPPLEMENTARY INFORMATION for

# Impact of refractive index heterogeneity on stimulated Brillouin scattering microscopy: a quantitative analysis

MENG XU,[1,2,†] ZIXUAN DU,[1,2,†] YUN QI,[1,2,*] JINRUI ZHANG,[1] SHUAI YAO,[1,2] ROBERT PREVEDEL,[3] AND FAN YANG,[1,2,*]

[1] Shanghai Institute of Optics and Fine Mechanics, Chinese Academy of Sciences, Shanghai, China
[2] University of Chinese Academy of Sciences, Beijing, China
[3] European Molecular Biology Laboratory, Heidelberg, Germany

[†]These authors contributed equally to this work.
[*]qiyun@siom.ac.cn;

[*]yang@siom.ac.cn

**Below SI file includes:**

Supplementary Notes 1-5
Figs. S1-S5

**Supplementary Note 1 Pulsed-SBS microscopy setup**

The experimental setup of the pulsed-SBS microscopy system is illustrated in Supplementary Fig. S1. Two 780 nm lasers serve as the pump and probe sources. To ensure high signal-to-noise ratio, the pump beam is modulated by an acousto-optic modulator (AOM1) to generate a pulse train (e.g., 40 ns pulse width). By converting CW beams into pulsed beams with high peak power but comparable average power, the pulsed scheme enhances the SBS signal by a factor inversely proportional to the duty cycle, while the shot-noise level remains unchanged, thereby improving the SNR. Crucially, this pulse train is further amplitude-modulated by a sinusoidal envelope provided by the internal reference source of the lock-in amplifier (LIA), which simultaneously serves as the reference for synchronous demodulation. This shifts the signal to a high-frequency band where the noise floor is significantly lower, further enhancing the SNR. The pump beam is then adjusted to p-polarization by a half-wave plate ($\lambda/2$), allowing it to pass through a polarizing beam splitter (PBS). After being converted to right-hand circular polarization by a quarter-wave($\lambda/4$) plate, it is focused into the sample by objective lens (Obj. 1, NA=0.7).

The probe beam frequency is scanned across the Brillouin gain peak by applying a sawtooth voltage waveform from a signal generator, with a period matched to the single-pixel dwell time, continuously sweeping the probe laser output frequency over a range of approximately 2–3 GHz. The frequency axis is calibrated in real time via a heterodyne beat-frequency method: the pump and probe seed beams are combined through a fiber coupler and detected by a broadband photodetector, with a frequency counter recording the instantaneous frequency difference between the two lasers. The probe beam is similarly pulsed by AOM2. A beam splitter (BS) directs a portion of the probe beam to photodetector (PD2) as a reference, while the remaining light is converted to left-hand circular polarization and focused into the sample via Obj. 2 (NA=0.7). The spatial overlap of the counter-propagating pump and probe beams induces stimulated Brillouin gain on the probe intensity.

After interaction, the signal-carrying probe beam is collected by Obj. 1, converted to s-polarization by the $\lambda/4$ plate, and reflected by the PBS toward PD1. A rubidium (Rb) cell is placed before the detector as a narrow-band notch filter to suppress residual pump leakage. Finally, the LIA performs differential detection between PD1 and PD2 to cancel common-mode noise, extracting the SBS signal amplitude. Three-dimensional SBS images are reconstructed by raster-scanning the sample with a motorized piezo stage and performing spectral fitting at each spatial coordinate.

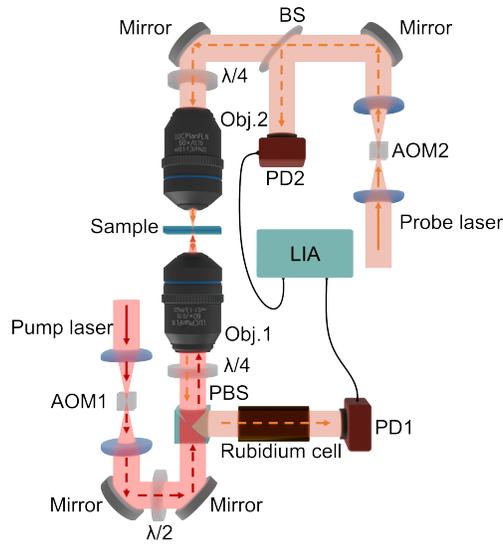

**Fig. S1.** Schematic of the SBS microscopy system. AOM: acousto-optic modulator, BS: beam splitter, LIA: lock-in amplifier, Obj: objective lens, PBS: polarization beam splitter, PD: Photodiode, λ/2: half-wave plate, λ/4: quarter-wave plate.

## Supplementary Note 2 Simulation model for pump–probe beam overlap

To quantitatively evaluate the effect of RI heterogeneity on the spatial overlap of pump and probe beams, a three-dimensional electromagnetic field propagation model was established using COMSOL Multiphysics. The model solves Maxwell's equations to obtain the spatial intensity distributions $I_p(\mathbf{r})$ and $I_s(\mathbf{r})$ of the pump and probe beams in the focal region, respectively.

Geometry and materials: The computational domain covers ±10 μm laterally and ±18 μm axially around the focal point, which is sufficient to fully encompass the focal field distribution of an NA = 0.7 focused beam. The sample consists of a PDMS bead (diameter ~8 μm, $n_{PDMS}$ = 1.41) embedded in an agarose gel matrix ($n_{gel}$ = 1.33), as shown in Fig. S2. Since the incident light is linearly polarized and the parametric scan of the microsphere position is confined to the xz-plane, the system possesses mirror symmetry about the yz-plane. A symmetry boundary condition is therefore applied, reducing the number of degrees of freedom to approximately half of the full model.

Electromagnetic solver: The beam envelope method in COMSOL is employed, which decomposes the electric field into a rapidly oscillating carrier and a slowly varying envelope, solving only for the envelope. This relaxes the mesh size requirement from λ/6 (as needed by full-wave methods) to nearly wavelength-scale, significantly reducing computational cost while maintaining accuracy. Both pump and probe beams operate at λ = 780 nm and are modeled as counter-propagating Gaussian beams incident from the +z and −z boundaries, respectively, simulating the counter-propagating focusing configuration with NA = 0.7 objectives. We note that the Gaussian beam model is based on the paraxial approximation, which becomes less accurate at high NA, resulting in a slightly larger simulated focal spot compared to the true diffraction limit. Matched boundary conditions are applied at the exit boundaries to absorb outgoing electromagnetic waves and suppress boundary reflections. Compared to perfectly matched layers (PML), matched boundary conditions require no additional absorbing layer thickness, thereby further reducing the total number of degrees of freedom.

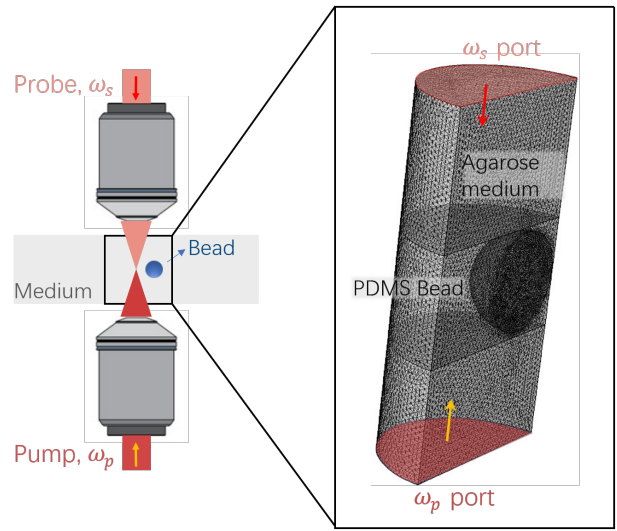

**Fig. S2 Numerical simulation setup in COMSOL Multiphysics.** The left panel shows the experimental configuration of counter-propagating pump and probe beams. The right panel displays the 3D computational domain featuring the PDMS bead embedded in agarose medium, with tetrahedral meshing applied for electromagnetic field calculation.

Meshing: Free tetrahedral meshes are used with a domain-specific refinement strategy. In the agarose gel region, the maximum element size is set to 0.4 μm. Within and near the PDMS microsphere, partial reflections at the RI interface create more complex field structures, requiring finer spatial sampling; the maximum element size is therefore refined to 0.25 μm (Fig. S2).

Parametric scanning: To simulate the point-by-point scanning process in experiments, the microsphere center position $(x_0, z_0)$ is parametrically scanned in the xz-plane while keeping the beam configuration fixed. This approach is adopted because the Gaussian beam analytical solutions used in the matched boundary conditions are predefined with respect to the boundary-to-waist distance. Fixing the beams and moving the microsphere avoids paraxial approximation errors that would arise from shifting the beam waist position, and is physically equivalent to the sample translation performed by the piezo stage in experiments.

Post-processing: For each microsphere position $(x_0, z_0)$, the volume integrals of $\kappa I_p(x, y, z) I_s(x, y, z)$ are computed separately over the PDMS bead and the agarose gel regions using COMSOL's derived values functionality. The integration domain is confined to a finite region around the focal point that contains the vast majority of the optical field energy; enlarging this domain has negligible effect on the results. This yields the data presented in Fig. 2(a) and (c).

### Supplementary Note 3 Justification of the polarization choice

Linear polarization is adopted for all simulations in this work. Due to computational memory limitations, symmetric boundary conditions are employed to reduce memory usage. These boundary conditions constrain the electric field vector to be either perpendicular or parallel to the symmetry plane, thereby precluding the use of circularly polarized light as employed in the experiments. To verify the validity of this choice, we computed the field distributions for both p-polarization and s-polarization separately, and found negligible differences under the current beam spot size configuration, as shown in Fig. S3. Therefore, the use of linearly polarized light does not affect the conclusions of this work.

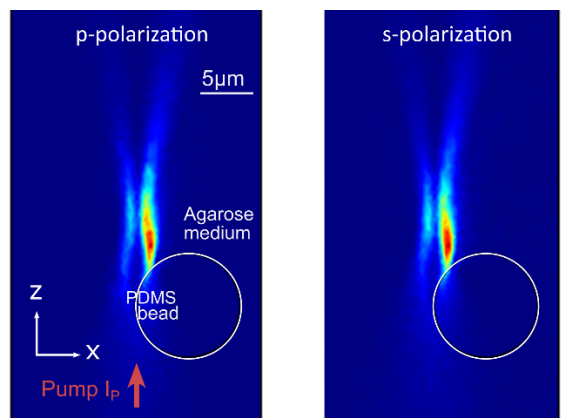

**Fig. S3 Field distributions for (a) p-polarization (b) and s-polarization.**

**Supplementary Note 4 Simulation model for fiber coupling efficiency**

To examine whether the fiber-coupled signal can serve as a faithful proxy for the SBS gain, a two-dimensional simplified model was constructed in COMSOL Multiphysics to capture the essential physics. The model features a mirror-symmetric dual-fiber, dual-lens configuration with a bead (radius = 3 μm) at the center. A schematic of the model geometry is shown in Fig. S4 (note that the fiber sections are truncated in the illustration for clarity; the actual computational domain extends to 300 μm along the propagation direction on each side).

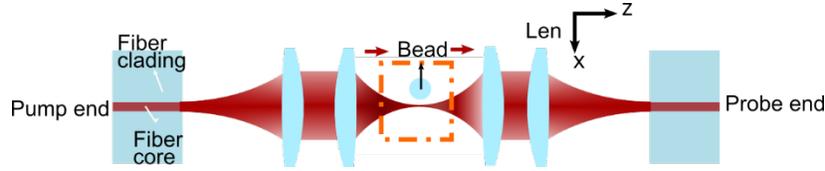

**Fig. S4 Schematic of the two-dimensional model.**

Fiber section: Located at the far left (and mirrored at the far right), with a total height of 52 μm. The core diameter is set to approximately four times the working wavelength (~3.1 μm), with $n_{core}$ = 1.51 and $n_{clad}$ = 1.50. The fiber length along the propagation direction is 300 μm, with 1-μm-thick PML regions at the top and bottom to absorb radiation modes. Numeric ports are defined at both fiber ends, with the fundamental mode profile pre-computed via boundary mode analysis. At the input port, wave excitation is enabled to launch the fundamental mode as the incident field; at the output port, wave excitation is disabled, allowing this port to serve as a passive receiver for evaluating the fiber coupling efficiency $\eta_c$.

Lens section: Positioned between the fiber and the sample. Due to the dimensional constraints of the 2D model, a free-form micro-lens (diameter 50 μm, $n_{lens}$ = 1.7) was used solely to achieve diffraction-limited focusing with a single element, without considering practical engineering constraints. This avoids the additional interface reflections and computational complexity associated with multi-element lens systems. The working distance is set to 25 μm, providing sufficient clearance to accommodate the full range of bead scanning positions in subsequent parametric sweeps. The lens surface profile is exported as coordinate data and imported into COMSOL to reconstruct the geometric boundary. A rigorous NA calibration was not performed, as the axial resolution in 2D and 3D models differs fundamentally due to energy conservation constraints — 2D focused beams exhibit slower axial decay than their 3D counterparts, making precise NA matching of limited practical significance. The primary purpose of this model is to examine the differential response between $\eta_c$ and the overlap integral, rather than to reproduce the absolute resolution of the experimental system.

The complete dual-side optical path is constructed by mirroring the single-side model about the sample center, with a bead (RI contrast of approximately 0.1 relative to the surrounding medium) placed at the center.

Boundary conditions: Transition boundary conditions are applied at the fiber–air and air–lens interfaces to suppress Fresnel reflections. The beam envelope method is again employed for the electromagnetic solver. PML is applied at the output-side fiber end to prevent back-reflections. The parametric scanning strategy is identical to Model 1: the optical path is fixed while the microsphere center is scanned in the $xz$-plane.

Solution procedure: (i) Boundary mode analysis is performed to compute the fundamental mode profile at the fiber ports (executed once and reused throughout the parametric scan). (ii) The computed fundamental mode is then used as the input field at the fiber excitation port. For each microsphere position $(x, z)$, the steady-state electromagnetic field distribution is solved across the entire computational domain. (iii) In post-processing, two key quantities are extracted: the overlap integral $\Psi(x_0, z_0) = \iint I_p(x,z) \, I_s(x,z) \, dS$ computed separately for each material in the focal region, and the fiber coupling efficiency $\eta_c(x_0, z_0)$ evaluated by COMSOL based on the modal overlap with the pre-computed fundamental mode at the output port. This yields the two-dimensional spatial distributions of $\Psi$ and $\eta_c$ in the $xz$-plane.

## Supplementary Note 5 Quantitative analysis of $A_{SBS}$ and $\eta_c$

To quantify the discrepancy between the norm. $A_{SBS}$ and norm. $\eta_c$, we define the relative difference ($\delta$) as:

$$\delta = \frac{\text{norm.} A_{SBS} - \text{norm.} \eta_c}{\text{norm.} A_{SBS}} \tag{S1}$$

$\delta = 0$ indicates that norm. $A_{SBS}$ and norm. $\eta_c$ are consistent, whereas $\delta = 1$ indicates that norm. $\eta_c$ is much lower than norm. $A_{SBS}$.

In Figs. 4(e) and 4(f), a white mask highlights a critical region spanning both bead-affected and unaffected signals for experimental and numerical data, respectively. The corresponding δ is quantified in Supplementary Figs. S5(c) and S5(b), where the maps illustrate the spatial distribution of this divergence across the sample interface. Both simulation and experiment demonstrate a consistent trend: δ surges toward 1 in regions where the focus field is distorted by the bead, while remaining near zero in unaffected areas. Crucially, this disparity is most pronounced in the regions where signal compensation is most essential, confirming the fundamental insufficiency of using fiber-coupling efficiency as a reliable proxy for Brillouin gain in the presence of focus field distortion by heterogeneous RI distributions. The relatively smaller spatial extent of the δ~1 region in the simulation is attributed to the simplified model employed for computational efficiency. Since the simplified model yields a larger size of focus field compared to the experimental focus, the area significantly influenced by RI heterogeneity is effectively reduced.

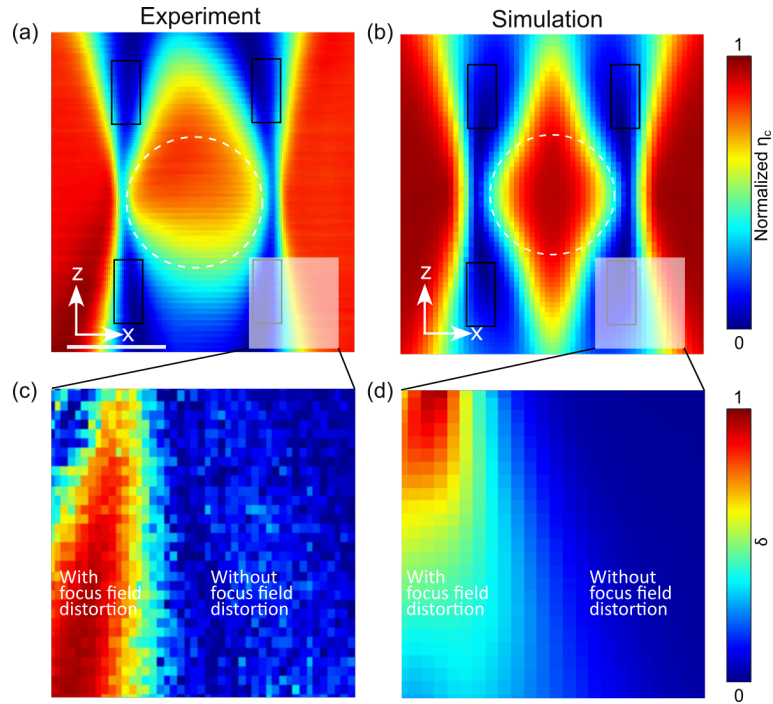

**Fig. S5.** (a), (b) Measured and simulated XZ norm. $\eta_c$ maps (same as Figs. 4(e) and 4(f)). Quantify the discrepancy between the norm. $A_{SBS}$ and norm. $\eta_c$ White mask: region of interest spanning both bead-affected and unaffected regions for the divergence between $A_{SBS}$ and $\eta_c$ analysis. (c) Experimental results and (d) numerical simulations of the spatial distribution of δ. This quantitative analysis corresponds to the region highlighted by the white mask in(a) and (b).